# Early diagnosis of Alzheimer's disease from MRI images with deep learning model

Sajjad Aghasi Javid
*Faculty of Electrical and Computer Engineering*
*University of Tabriz*
Tabriz, Iran
s.aghasi1400@ms.tabrizu.ac.ir

Mahmood Mohassel Feghhi
*Faculty of Electrical and Computer Engineering*
*University of Tabriz*
Tabriz, Iran
mohasselfeghhi@tabrizu.ac.ir

*Abstract*— It is acknowledged that the most common cause of dementia worldwide is Alzheimer's disease (AD). This condition progresses in severity from mild to severe and interferes with people's everyday routines. Early diagnosis plays a critical role in patient care and clinical trials. Convolutional neural networks (CNN) are used to create a framework for identifying specific disease features from MRI scans Classification of dementia involves approaches such as medical history review, neuropsychological tests, and magnetic resonance imaging (MRI). However, the image dataset obtained from Kaggle faces a significant issue of class imbalance, which requires equal distribution of samples from each class to address. In this article, to address this imbalance, the Synthetic Minority Oversampling Technique (SMOTE) is utilized. Furthermore, a pre-trained convolutional neural network has been applied to the DEMNET dementia network to extract key features from AD images. The proposed model achieved an impressive accuracy of 98.67%.

## I. Introduction

AD calls for intensive medical attention. Early and accurate diagnosis is crucial to initiate the clinical process of AD and halt the progression of the disease. AD has a negative impact on the brain's normal functioning, affecting daily tasks such as speaking, writing, and reading [1]. AD is a neurological condition that damages brain tissue over time. It accelerates the loss of even the most basic skills by causing memory and cognitive deficiencies. Physicians classify Alzheimer's disease (AD) using computer-aided diagnostic techniques and neuroimaging in its early stages. As to the latest census report of the AD Association, this condition affects over 9.4 million Americans who are 65 years of age or older [2]. AD association estimates that in the next 50 years, 60 million people may develop dementia. Between 60% and 80% of dementia cases globally are thought to be caused by AD. Every 3 seconds, someone in the world develops dementia, with a 60% likelihood of it being AD [3]. The early diagnosis and automatic classification of AD represent a burgeoning field that has resulted in new insights in large-scale multimodal neuroimaging. Various methods, such as MRI, Positron Emission Tomography (PET), and genetic sequencing, are utilized in AD research. One major drawback of these methods is their time-consuming nature. Moreover, patients are exposed to the effects of radioactivity. This article focuses on the use of MRI images due to their advantages, enhanced flexibility in imaging, superior tissue contrast, low ionizing radiation, and the capacity to offer insightful information on the structure of the human brain [4]. Pre-trained models have shown superior performance in automatically identifying stages of AD from MRI scans. These models are trained on extensive datasets, enabling them to deliver improved results [5]. In the proposed model of this article, the initial issue of imbalanced data obtained from the Kaggle database [6] is tackled by employing SMOTE technique to balance the dataset. In the next step, key features of AD are extracted from brain MRI images using a pre-trained CNN. Since the convolutional core of pre-trained models like Inceptionv3 excels in extracting local features, integrating the output of Inceptionv3 with a low-parameter Alzheimer's network such as DEMNET results in improved accuracy for AD diagnosis. The innovation of this paper lies in the combination of the pre-trained Inceptionv3 model for extracting local features with the Alzheimer's-specific DEMNET network. Based on our research so far, the application of the Inceptionv3 model to MRI images and the fusion of its extracted features with the DEMNET neural network have not been previously explored.

## II. RELATED WORKS

Deep learning has garnered significant attention for its potential in the diagnosis of AD. Several recent deep learning approaches have been proposed as tools to aid physicians in making informed medical decisions regarding this disease. Here, we present a few noteworthy studies in the subject.

Lu et al. [7] created a multi-stage deep neural network with an 82% accuracy rate to identify people with moderate cognitive impairment who are at risk of acquiring AD within three years. Their model's accuracy was 83% in cases without dementia and 94% sensitive in cases of AD. Gupta et al. [8] demonstrated a diagnostic method for classifying AD utilizing integrated features from MRI images' cortical, subcortical, and hippocampus areas. This study is based on data from The Alzheimer's Disease Neuroimaging Initiative (ADNI). The goal of the long-term, multicenter ADNI project is to create biochemical, genetic, imaging, and clinical biomarkers for AD that can be used for tracking and early diagnosis. This method enhanced the accuracy of 96.5% in separating AD patients from healthy people. Ahmed et al. [9] proposed a feature extraction and classification model that uses the regions of the left and right hippocampus regions in MRI scans to prevent overfitting and obtains a 90% accuracy rate for the diagnosis of AD. Basher et al. [10] introduced a method for automatically segmenting regions of interest from large MRI volumes for AD diagnosis, achieving accuracies of 94.82% and 94.02% based on the left and right hippocampi, respectively. Nawaz et al. [11] revealed a pre-trained Alexnet model It resolved class imbalance and attained 99% classification accuracy using random forest (RF), k-nearest neighbors (KNN), and support vector machines (SVM). Ieracitano et al. [12] suggested a data-driven approach to identify AD utilizing brain wave analysis and decomposition. A CNN was then used for classification, and the results showed 89% accuracy for binary classification and 83% accuracy for multi-class classification from 2D images. Jain et al. [13] utilized a pre-trained VGG16 model for feature extraction, FreeSurfer for preprocessing, entropy-based slice selection, and transfer learning using the PFSECTL mathematical model for MRI-based AD classification, achieving an accuracy of 95%. Mehmood et al. [14] used tissue segmentation to extract gray matter for AD classification, achieving a classification accuracy of 98% for AD patients compared to healthy individuals. Shi et al. [15] presented a deep polynomial network for the identification of AD that performed well on both small and large datasets, utilizing AD datasets to achieve 55.34% accuracy for binary and multi-class classification. Liu et al. [16] investigated the discriminating potential of whole-brain volume asymmetry using a Siamese neural network. The group combined a novel non-linear kernel technique with cloud-based MRI processing to generate low-dimensional volumetric information for pre-defined brain structure atlases. The suggested network attained a 92% balanced accuracy. Afzal et al. [17] addressed the class imbalance problem in AD detection using a data augmentation framework and achieved a classification accuracy of 98.41% in a single view and 95.11% in 3D view of the OASIS dataset. Fu'adah et al. [18] When employing the Alexnet architecture and MRI datasets, it is feasible to attain a maximum accuracy of 95% and a loss of 0.1643% for distinguishing between non-demented, very mild, and moderately severe dementing states. The training process utilizes the Adam optimizer with a range of learning rates, including 0.0001, 0.001, 0.01, and 0.1, along with the binary cross-entropy loss function.

## III. PROPOSED WORK

Deep learning has received a lot of interest lately in a variety of domains, including brain imaging, energy management systems, cervical cancer diagnosis, and malaria detection [19]. The proposed method in this paper involves four main stages: data preprocessing, dataset balancing using the SMOTE technique, feature extraction from MRI images using the pre-trained Inceptionv3 model, and final image classification using the DEMNET model. Each stage of the proposed model will be elaborated on in the following sections. Despite the literature's abundance of machine learning and deep learning approaches for AD classification, multiclass AD classification's class imbalance and large number of model parameters remain major obstacles. Through the process of linking a data point with its k-nearest neighbors, the SMOTE method creates samples of a given class. Synthetic data points produced by the SMOTE technique are not an exact duplicate of the minority class case.

### A. Dataset Description

Visual examples of the four classes of images are presented in Fig. 1.

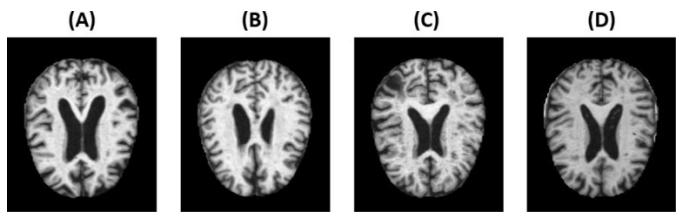

Figure 1 (A) MID (B) MOD (C) ND (D) VMD.

The AD dataset was obtained via Kaggle. Kaggle is open-access platform. 6400 MRI scans divided into four categories are available on Kaggle.: Mild Demented (MID), Moderate Demented (MOD), Non-Demented (ND), and Very Mild Demented (VMD). Images in the collection are first downsized to 128 × 128 from dimensions of 208 × 176. Table 1 presents the dataset breakdown, highlighting its class imbalance. To address this issue, the SMOTE technique is employed .SMOTE involves randomly replicating samples from the minority classes to match the majority class. In this study, a random seed of 42 is used for the SMOTE technique. One of the benefits of using SMOTE is its ability to reduce overfitting. Table 2 displays the dataset distribution after augmentation using SMOTE, resulting in a total of 12800 images, with 3200 images in each class.

Table 1 Number of data in each class before SMOTE.

| Class | No of Images |
|---|---|
| Mild Demented | 896 |
| Moderate Demented | 3200 |
| Non-Demented | 2240 |
| Very Mild Demented | 64 |

Table 2 Number of data in each class after SMOTE.

| Class | No of Images |
|---|---|
| Mild Demented | 3200 |
| Moderate Demented | 3200 |
| Non-Demented | 3200 |
| Very Mild Demented | 3200 |

The dataset consists of 12800 samples, which are divided into 80% for training about 10240 samples, 10% for validation, and 10% for testing.

## B. Feature extraction using InceptionV3

Recent research in medical imaging has demonstrated the superior performance of pre-trained deep convolutional networks, particularly those trained on large datasets like ImageNet, for classifying medical images [20]. Medical datasets are often challenging to acquire and are typically small in size, posing difficulties for CNNs and increasing the risk of overfitting. Pre-trained models offer a viable solution to these challenges [21]. In this study, we employ the Inceptionv3 pre-trained deep convolutional network [22]. Inceptionv3 represents the third iteration of the Inception architecture and boasts several improvements over its predecessors, including factorized convolutions that cut down on parameters without sacrificing network efficiency. The DEMNET network receives its input as the output of the Inceptionv3 network., as depicted in Fig. 2. To extract features from MRI images using Inceptionv3, we remove the softmax layer of the network. this allowing us to directly obtain feature maps from the preceding layer. The optimization technique used to train the proposed model was the Root Mean Square Propagation (RMS prop) optimizer. The RMS prop was created specially to manage chaotic mini-batch learning. The gradient is normalized using the moving average of square gradients in order to resolve the backpropagation problem with the vanishing gradient. In the optimization phase, this normalization aids in maintaining momentum balance. The adaptive learning rate used by RMS prop modifies over time based on the accumulated square of previous gradients, as opposed to being treated as a constant hyperparameter. This adaptive nature allows RMS prop to adjust the learning rate according to the specific requirements of the optimization process.

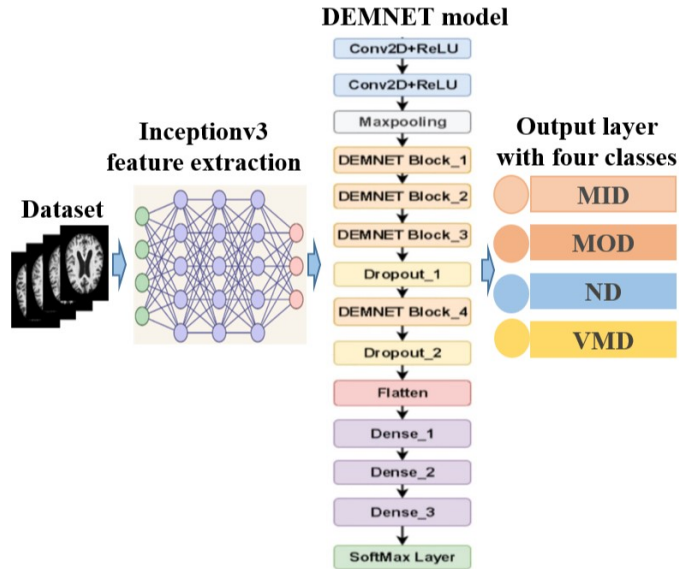

Figure 2 Proposed hybrid model framework with feature extraction by pre-trained CNN with Inceptinv3.

## C. DEMentia NETwork (DEMNET)

After extracting the essential features, the MRI images are input into a convolutional neural network CNN. The DEMNET architecture [23] contains four DEMNET blocks, two dropout layers, three dense layers, one Softmax classification layer, two convolutional layers with the Rectified Linear Unit (ReLU) activation function, and one Max-pooling layer. The model's convolutional layers are essential because they take an image as input and mix it with weight filters to extract features. To minimize the image's computational and geographical dimensions, the pooling layers are used in between the convolutional layers. The maximum pooling operation is calculated using Equation (1) in the following manner.:

$$\text{MP} = \text{Floor}\left(\frac{I_x - p}{S} + 1\right) \quad (1)$$

$I_x$ is the input form, $P$ is the pooling window size, and $S$ is the stride.

The DEMNET block is composed of a batch normalization layer, a max-pooling layer, and two stacked convolutional layers that use the ReLU activation function. It utilizes different filters, including 32, 64, 128, and 256, to classify the stages of AD. This design can effectively highlight each stage of cognitive decline in an image. In the ReLU activation function, if the input neuron is positive, it retains its value. otherwise, it is replaced by zero as shown in equation (2).

$$F = (0, x) \quad (2)$$

The value of $x$ in the equation is the input of the neuron. Following the convolutional layers, batch normalization is applied. The DEMNET block uses batch normalization, a regularization technique, to reduce overfitting problems. Because the number of classes is 4, the number of neurons in the softmax function is equal to 4 [24].

## IV. EVALUATION METRICS

Several metrics are employed to assess the effectiveness of the proposed model, including recall, accuracy, precision, and F1 score, which is obtained from the confusion matrix [25]. A summary of the model's performance is given by this matrix [26]. Equation (3) can be used to determine accuracy, a critical statistic for assessing the model's predictive power for both positive and negative examples [27].

$$\text{Accuracy} = \left(\frac{TP + TN}{FP + TP + TN + FN}\right) \quad (3)$$

The number of times the system accurately calls those who have the disease True Positives (TP). Similarly, the count of samples of healthy individuals that the system correctly identifies as True Negatives (TN) is noted. False Positives (FP) are instances where the system incorrectly identifies healthy samples as diseased, while False Negatives (FN) are instances where the system incorrectly identifies diseased samples as healthy [28]. The ratio of the actual positive observation to the total positive forecast is known as precision. A high accuracy, approaching 1, indicates strong performance of the proposed model [29]. Precision can be computed using equation (4).

$$\text{Precision} = \left(\frac{TP}{TP + FP}\right) \quad (4)$$

The Recall metric can be described as sensitivity, representing the classifier's capability to accurately locate all positive instances. It can be computed using equation (5).

$$\text{Recall} = \left(\frac{TP}{TP + FN}\right) \quad (5)$$

According to equation (6), The average harmonic of recall and precision is known as the F1-score.

$$\text{F1-Score} = \left(\frac{2 * Precision * Recall}{Precision * Recall}\right) \quad (6)$$

## V. RESULTS AND DISCUSSIONS

This article's model was trained with the gradient descent optimization approach over 50 epochs with a batch size of 128 and an initial learning rate of 0.001. Employing the SMOTE technique, the model achieves an impressive detection accuracy of around 98.67%. The model's evaluation using the test dataset produces a confusion matrix. Fig. 3 shows the confusion matrix of the proposed model for classifying stages of cognitive decline to predict AD. This matrix compares predicted classes with labeled classes across four different categories. It reflects the model's performance on the training dataset and is calculated from 302 images of MID, 335 images of MOD, 301 images of ND and 342 images of VMD. Table 3 presents the evaluation metrics for each class based on the confusion matrix.

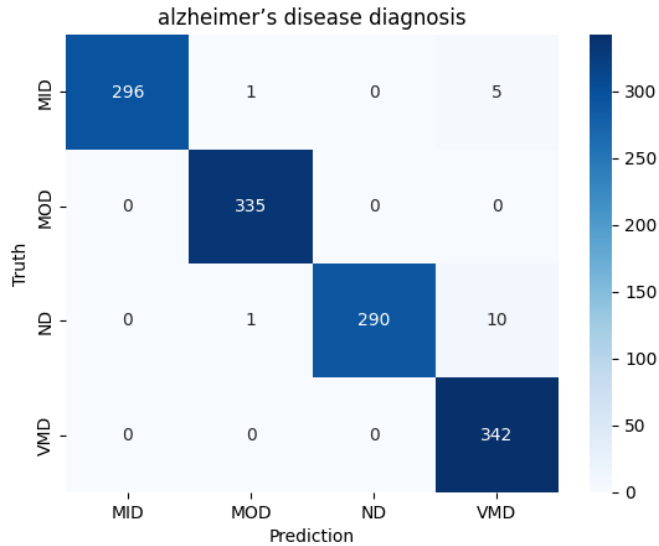

Figure 3 Confusion matrix obtained from 4 classes for the proposed model of this paper.

Table 3 Performance Measures for 4 Class.

| Diseases | Precision | Recall | F1-score |
| --- | --- | --- | --- |
| MID | 1.00 | 0.98 | 0.99 |
| MOD | 0.99 | 1.00 | 1.00 |
| ND | 1.00 | 0.96 | 0.98 |
| VMD | 0.96 | 1.00 | 0.98 |

## VI. CONCLUSION

AD a progressive neurological disorder, gradually impairs memory and basic cognitive functions. It is one of the most common forms of cognitive decline in older adults. In its early stages, AD affects areas like the hippocampus and entorhinal cortex, crucial for memory. As it advances, it impacts broader brain regions, leading to significant challenges in social interactions, language, and thinking abilities. Despite the fact that AD cannot currently be treated, early detection and the right medicine can help moderate the disease's progression. Scholars worldwide have looked into deep learning and machine learning techniques for AD early detection. This article presents a hybrid model that merges the capabilities of the DEMNET neural network with those of Inceptionv3 to extract comprehensive features from MRI images. By leveraging the strengths of both models, this hybrid approach aims to capture nuanced patterns and representations that are crucial for accurate Alzheimer's disease classification. The study uses brain MRI images across four classes: mild cognitive impairment, moderate cognitive impairment, noncognitive impairment, and very mild cognitive impairment. The proposed model achieves a 98.67% accuracy in classifying AD stages, outperforms the other evaluated models.